\documentstyle[preprint,prl,aps,epsf]{revtex}
\begin{document}

\draft

\title{
On the Bilayer Coupling in the Yttrium-Barium Family
of High Temperature Superconductors
}

\author{A. J. Millis $^a$ and H. Monien $^b$}
\address{
$^a$
AT\&T Bell Laboratories\\
600 Mountain Avenue\\
Murray Hill, NJ 07974\\
$^b$
Theoretische Physik\\
ETH Z\"urich\\
CH-8093 Z\"urich\\
Switzerland
}

\date{June 16, 1995}

\maketitle

\begin{abstract}
We present and solve a model for the susceptibility of two CuO$_2$
planes coupled by an interplane coupling $J_\perp$ and use the results to
analyze a recent "cross-relaxation" NMR experiment on
Y$_2$Ba$_4$Cu$_7$O$_{15}$.  We deduce that in this material the
product of $J_\perp$ and the maximum value of the in-plane susceptibility
$\chi_{max}$ varies from approximately 0.2 at $T$ = 200~K to 0.4 at
$T$ = 120~K and
that this implies the existence of a temperature dependent
in-plane spin
correlation length.  Using estimates of $\chi_{max}$ from the literature
we find 5~meV $< J_\perp <$ 20~meV.
We discuss the relation of the NMR results to
neutron scattering results which have been claimed to imply that in
YBa$_2$Cu$_3$O$_{6+\rm x}$ the two planes of a bilayer are perfectly
anticorrelated.  We also propose that the recently observed 41~meV
excitation in YBa$_2$Cu$_3$O$_7$ is an exciton pulled down below the
superconducting gap by $J_\perp$.
\end{abstract}

\vfill\eject

In the yttrium-barium (Y-Ba)
family of high temperature superconductors  the basic structural
unit is a "bilayer", consisting of two CuO$_2$ planes; the bilayers
are separated by CuO chains.  Neutron scattering \cite{Tranquada92}
and more recently NMR experiments \cite{Stern95} have shown that the
Cu spins on adjacent planes in a bilayer are coupled.  Intra-bilayer
coupling has been shown theoretically to lead to a "spin gap"
\cite{Altshuler92,Millis93,Altshuler94,Ubbens94} similar to that observed
\cite{Takigawa91} in NMR experiments on underdoped members of the YBa
family.  In view of the great importance of the spin gap phenomenon, a
quantitative analysis of the spin dynamics of a bilayer is desirable.
In this letter we provide this analysis and use the results to
interpret NMR and neutron scattering experiments.

Focus on the two planes of a bilayer, and neglect coupling to other
bilayers.  We label the spin degrees of freedom by an index
$a$ = 1, 2 distinguishing planes, and a site index $i$. We then define the
susceptibility
\begin{equation}
   \chi^{ab}(q,\omega) =
   \int_{0}^{\infty}dt\;
   e^{i\omega t + \vec q\cdot\left(\vec{R}_i-\vec{R}_j\right)}
   \left<\left[S_i^a(t), S_j^b(0)\right]\right>.
\end{equation}
Because of the symmetry under exchange of planes, $\chi^{ab}$ has only
two independent components, $\chi^{11}(q,\omega)=\chi^{22}(q,\omega)$
and $\chi^{12}(q,\omega)=\chi^{21}(q,\omega)$.  The two independent
components of $\chi$ may be taken to be the even and odd (under the
interchange of planes) components $\chi_{even,
odd}=\chi_{11}\pm\chi_{12}$.  In a system with antiferromagnetic
coupling one expects that at large wavevectors the odd-parity spin
fluctuations are softer than the even parity spin fluctuations.

Further analysis requires a model.  We shall assume that the
interplane coupling $J_\perp$ is weak in comparison to the energies
determining the spin susceptibility $\chi_0(q,\omega)$ of a single
plane and that its effects may be modeled via the
RPA. Thus we write
\begin{equation}
        \bbox{\chi}^{-1} = \left[
        \begin{array}{cc}
                \chi^{-1}_{0}(q,\omega) & -J_\perp  \\
                -J_\perp & \chi^{-1}_{0}(q,\omega)
        \end{array}
        \right].
        \label{eq:chi-rpa}
\end{equation}
We further assume $\chi_{0}(q,\omega)$ has the scaling form
$\chi_{0}(q,\omega) = \chi_0 \xi^z f(q\xi,\omega/\xi^z)$, where $f$ is
a scaling function normalized so that $f(0,0)=1$, $z$ is the dynamical
exponent, $\xi$ is a correlation length and $\vec q$ is measured from
an ordering wavevector $\vec Q$ which for the present discussion is
arbitrary. $\vec Q$ is believed to be of the order of $(\pi,\pi)$ in
high T$_{\rm c}$ materials.
{}From Eq. \ref{eq:chi-rpa} we see that
$\chi^{11}(q,\omega) =
\chi_0(q,\omega)/(1-(J_\perp\chi_0(q,\omega))^2)$ and
$\chi^{12}(q,\omega) = J_\perp
\chi_0(q,\omega)^2/(1-(J_\perp\chi_0(q,\omega))^2)$.  The crucial
parameter controlling the susceptibilities in the static limit is
\begin{equation}
        \Delta=J_\perp \chi_{max} =J_\perp\chi_0\xi^2.
        \label{eq:Delta}
\end{equation}
We must assume $\Delta<1$ so that the material has no long range
order.  If $\Delta^2 \ll 1$ then $\chi^{11} \approx \chi_0$ and
$\chi^{12} \approx J_\perp \chi_0^2$.  In this limit the interplane
coupling has a weak effect and the RPA is an appropriate model.  On
the other hand, if $\Delta ^2 \approx 1$ then the interplane coupling
is strong and the use of the RPA may be questioned.

We now turn to the NMR experiments of interest.  These are $T_2$
experiments performed on Y$_2$Ba$_4$Cu$_7$O$_{15}$, a material in
which the single-chain structure of YBa$_2$Cu$_3$O$_7$ alternates with
the double-chain structure of YBa$_2$Cu$_4$O$_8$ \cite{Stern94}.  As a
result, atoms on different planes of a bilayer have somewhat different
local environments and therefore somewhat different NMR resonance
frequencies, which may be independently studied. Despite the
differences in local environment the electronic properties of the two planes
are not very different \cite{Stern94}, so it is still appropriate to
model the eletronic properties with Eqs. (1-5). Now the NMR $T_2$
measures the rate at which a nuclear spin is depolarized by
interacting with other nuclear spins, i.e. it measures the
nuclear-spin--nuclear-spin interaction strength.  In high $T_c$
materials the dominant contribution to the nuclear-spin nuclear-spin
interaction comes from polarization of electronic spins, and may be
related to the static limit of the real part of the electronic spin
susceptibility.  In Y$_2$Ba$_4$Cu$_7$O$_{15}$ it is possible to
measure $T_{2}$, the rate at which a spin in one plane is depolarized
by spins in the same plane, and $T_{2\perp}$, the rate at which a spin
in one plane is depolarized by spins in the other plane.  $T_{2}$ is
related to the electronic spin susceptibility by
\cite{Pennington91}
\begin{equation}
    \frac{1}{T_{2}}=
    \left[
         \sum_q\left[A_q^2 \chi^{11}_q\right]^2-
         \left(\sum_q A_q^2 \chi^{11}_q\right)^2
    \right]^{1/2}
    \label{eq:T2}
\end{equation}
while $T_{2\perp}$ is given by
\cite{Monien94}
\begin{equation}
        \frac{1}{T_{2\perp}} =
        \left[\sum_q\left[A_q^2 \chi^{12}\right]^2\right]^{1/2}.
        \label{eq:T2perp}
\end{equation}

We have calculated $T_{2}$ and $T_{2\perp}$ from Eqs.
(\ref{eq:chi-rpa},\ref{eq:T2},\ref{eq:T2perp}). The precise values
obtained depend upon the form chosen for $f(q\xi)$. We have used two
forms for $f(x) = f(q\xi,\omega=0)$: a Lorentzian, $f(x)=1/(1+x^2)$,
and a Gaussian, $f(x)=\exp(-\log(2) x^2)$ ( the $\log(2)$ is
introduced so $f(x=1)=1/2$).
We measure $\xi$ in units of the lattice constant, we set $\hbar=1$ and
assume that the hyperfine coupling can be approximated by its value at
$Q$, $A_Q$. We find
\begin{equation}
        \frac{1}{T_{2}} = A_Q^2 \chi_0 \xi g_{in}(\Delta)
        \label{eq:T2in-ans}
\end{equation}
and
\begin{equation}
        \frac{1}{T_{2\perp}} = A_Q^2 J_\perp \chi_0^2 \xi^3 g_{\perp}(\Delta)
        \label{eq:T2perp-ans}
\end{equation}
where
$g_{in}$ and $g_\perp$ are defined in terms of the function $f(x)$ via
\begin{equation}
  g^2_{in}(\Delta) = \frac{1}{2\pi} \left[
  \int_0^\infty dx\; \frac{f^2}{(1-\Delta^2 f^2)^2} - \frac{1}{2\pi\xi^2}
  \left( \int_0^\infty dx\; \frac{f^2}{1-\Delta^2 f^2} \right)^2
  \right]
\end{equation}
\begin{equation}
  g^2_\perp(\Delta) = \frac{1}{2\pi}
  \int_0^\infty dx\; \frac{f^4}{(1-\Delta^2 f^2)^2}.
\end{equation}
In writing Eqs. (8,9) we have assumed that the correlation length
is so long that lattice effects may be neglected. We have investigated
this issue by performing the exact integrals numerically. The  parameter
governing the size of the lattice effects is $(\pi \xi)^{-1}$; for
$\xi \ge 1$ we have found that they are negligible.

In Fig. 1 we present the calculated results for $T_{2}/T_{2\perp}$ $=\Delta
g_{\perp}(\Delta) /g_{in}(\Delta)$. From the experimental values
$T_{2}/T_{2\perp} =0.15$ at $T$ = 200 K and $T_{2}/T_{2\perp} =0.30$ at $T$ =
120 K \cite{Stern95} we obtain $\Delta \approx 0.2$ at $T$ = 200 K and $\Delta
\approx 0.4$ at $T=120K$.  Note that even at the lowest temperature we find
that the system is in the small $\Delta$ regime in which $T_2/T_{2\perp}$
is linear in $\Delta$, suggesting that the
between-planes coupling is sufficiently small that the RPA formula is
justified.

Now Y$_2$Ba$_4$Cu$_7$O$_{15}$ has a doping somewhere between
YBa$_2$Cu$_3$O$_7$ and YBa$_2$Cu$_3$O$_{6.7}$ \cite{Stern94}.  It
seems clear that the strength of the magnetic correlations is a
relatively rapid function of doping and increases as one moves from
$O_7$ to the insulator, while $J_{\perp}$ is unlikely to be a
sensitive function of doping.  It therefore seems very likely that for
dopings corresponding to YBa$_2$Cu$_3$O$_{6.7}$ and below,
$\chi_{max}$ is so large that $\Delta \approx 1$ and the planes are so
strongly coupled that the properties are not linear in $J_\perp$.

The magnitude of $J_\perp$ may be determined if $\chi_{max}$ is known
and conversely.  If the susceptibility were only weakly $q$-dependent
then the measured uniform susceptibility $\chi_{uniform} \approx$ 2
states/eV-Cu \cite{Walstedt92}
would provide a good estimate for $\chi_{max}$ and our
value $\Delta \approx 0.4$ would imply $J_\perp \approx$ 0.2 eV.  Such
a value is very difficult to justify on microscopic grounds because the
insulating antiferromagnetic parent compounds of the high $T_c$
superconductors have in-plane exchange constants $J_{in-plane} \approx
0.12$ eV and it is generally believed that $J_{\perp} \ll
J_{in-plane}$.  Therefore, we believe the cross-relaxation results
imply $\chi_{max} \gg \chi_{uniform}$.  A similar conclusion has
been drawn from an
analysis of the magnitude of the in-plane $T_{2}$
measured on YBa$_2$Cu$_3$O$_7$ \cite{Pennington91}, combined
with various assumptions about magnitudes of hyperfine couplings.  The
magnitude of the hyperfine coupling drops out of the present analysis.
The estimate $J_\perp \approx 10-20$~meV has been
obtained from band structure calculations \cite{Andersen94}, implying
$\chi_{max} \approx$ 40 states/eV-Cu.  The in-plane
$T_2$ experiment led to the value $\chi_{max} \approx$ 80
states/eV-Cu \cite{Pennington91} implying $J_{\perp}
\approx 5$~meV.

In summary, the cross-relaxation experiment shows that the real part
of the susceptibility at some non-zero $q$ is much larger than the
uniform susceptibility.
Now the temperature dependence of the $T_2$ rates must be
due to the
temperature dependence of this antiferromagnetic maximum.
Two scenarios have been proposed for the temperature dependence: in
the {\it antiferromagnetic scenario} the temperature dependent
quantity is the correlation length $\xi$.  In the {\it
generalized marginal fermi
liquid scenario} the temperature dependent quantity is the overall
amplitude $\bar{\chi}$ \cite{Si94}.  From
Eqs. (\ref{eq:T2in-ans},\ref{eq:T2perp-ans}) we see that in the regime
where $T_{2\perp}$ is linear in $\Delta$ the antiferromagnetic
scenario predicts $T_{2}^3/T_{2\perp}$ is temperature independent,
while the marginal fermi liquid scenario predicts
$T_{2}^2/T_{2\perp}$ is temperature independent.  The
experimentally determined ratios are plotted in Fig. 2 and are
more consistent with the antiferromagnetic scenario.

The imaginary parts of the two independent susceptibilities
$\chi_{even}$ and $\chi_{odd}$ are measurable via neutron scattering
because they have different dependences on $q_z$, the momentum
transverse to the CuO$_2$ planes \cite{Tranquada89}.  Neutron
scattering experiments have been performed on a variety of members of
the yttrium-barium family of high-T$_{\rm c}$ materials
\cite{Tranquada92,Tranquada89,Rossad-Mignod91,Mook93}.  The
experimental result is that {\it only} $\chi_{odd}$ is seen. At
frequencies less than 30~meV and temperatures less than room
temperature the even parity fluctuations are
claimed to be completely frozen out.
The theory of neutron scattering in high $T_c$ materials is presently
controversial.  There is no generally accepted model which correctly
accounts for the observed lineshapes and temperature dependences.  To
investigate the connection between the cross-relaxation experiments
and neutron scattering we have chosen to calculate the ratio of the
$q$-integrated even and odd parity susceptibilities.  This ratio is
insensitive to the precise details of the susceptibilities.
For definiteness we used the "MMP", dynamical exponent $z=2$ ansatz
$\chi_0(q,\omega) = {\bar \chi} /(\xi^{-2} + q^2 - i\omega/\Gamma)$.  Here
$\Gamma$ is a microscopic spin relaxation time.  The results depend on
$\Delta$ and on $\omega_{SF} = \Gamma/\xi^2$, which is the softest
spin fluctuation frequency of a single plane.  Of course $J_{\perp}$
will reduce this frequency for the odd parity channel and increase it
for the even channel.  Results are shown in Fig. 3 for several values
of $\Delta$.  We see that the relative weight of the even parity
fluctuations becomes small only for $\Delta > 0.5$.  We believe that
the neutron results, which seem to require a $\Delta > 0.5$, are not in
contradiction to our analysis of the cross-relaxation experiment,
which yielded a $\Delta \le 0.4$, because the strongest neutron
evidence for locked bilayers was obtained from a study of
YBa$_2$Cu$_3$O$_{6.5}$ \cite{Tranquada92}, which as we have previously
noted is  closer to the magnetic instability than
Y$_2$Ba$_4$Cu$_7$O$_{15}$, and therefore may be expected to have a larger
$\Delta$.

Another experiment in which the even parity fluctuations were not seen
at all was an observation of a rather sharp peak at an energy of 41
meV in the {\it superconducting state} of YBa$_2$Cu$_3$O$_7$.  In this
material there is some evidence of a spin fluctuation peak in the
normal state at a similar energy, but it is much broader in $q$ and
$\omega$.  We suggest that the 41~meV peak is an exciton pulled down
below the superconducting gap edge by the interplane coupling
$J_{\perp}$.  We note that YBa$_2$Cu$_3$O$_7$ has somewhat weaker
magnetic correlations than Y$_2$Ba$_4$Cu$_7$O$_{15}$, so one would
expect $\Delta < 0.4$ in the normal state.  Now in the superconducting
state, the presence of the gap implies that $\chi^{''}$ becomes very
small at energies less than the gap, and has a peak (the details of
which depend in a complicated way on the details of the
superconducting order parameter) at the gap edge.  This behavior of
$\chi^{''}$ implies via the Kramers-Kronig relation a large
enhancement of the real part, $\chi^{'}$, at the gap edge.  We propose
that this enhanced $\chi^{'}$ leads to a weakly damped pole, at a
frequency of order the gap edge, in the odd parity channel of the RPA
formula, Eq.
\ref{eq:chi-rpa}, and that this pole produces the feature seen
experimentally.  Our proposal provides a natural explanation for the
sharpness of the feature and for the nearly perfect bilayer correlation
observed at low temperatures. If this scenario is correct the even
parity component of the neutron scattering signal should appear
above T$_{\rm c}$.

In summary, our analysis of the cross-relaxation
experiment  suggests that the between-planes
coupling $J_{\perp}$ is small, but has non-negligible effects which
furthermore increase as the temperature is decreased.  These are precisely
the assumptions made in the theories which attribute spin gap formation
to interplane pairing, so we believe our results tend to support this picture.

We thank T. M. Rice,
R. Stern and M. Mali for helpful discussions.
H. M. acknowledges the
hospitality of AT\&T Bell Laboratories where part of this
work was done.

\newpage

\begin{figure}[t]

\protect\centerline{\epsfbox{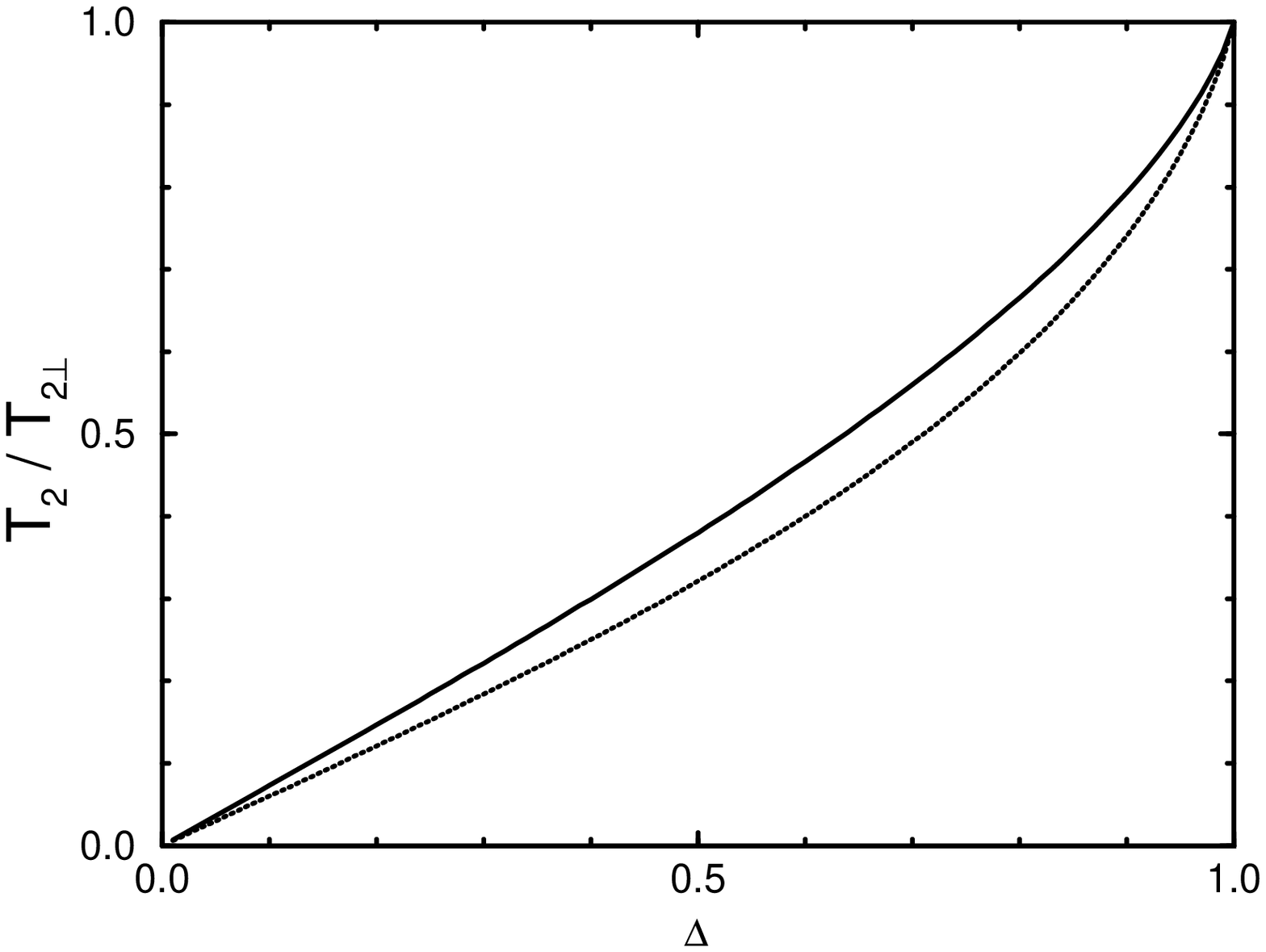}}

\protect\caption{ Ratio of cross-relaxation rate $1/T_{2\perp}$ to in-plane
  relaxation rate $1/T_{2}$ plotted versus coupling parameter $\Delta =
  J_{\perp}\chi_{max}$ for Lorentzian (dotted line) and Gaussian (solid line)
  form factors and calculated from Eqs. (6-9).
  }

\end{figure}

\begin{figure}[h]

\protect\centerline{\epsfbox{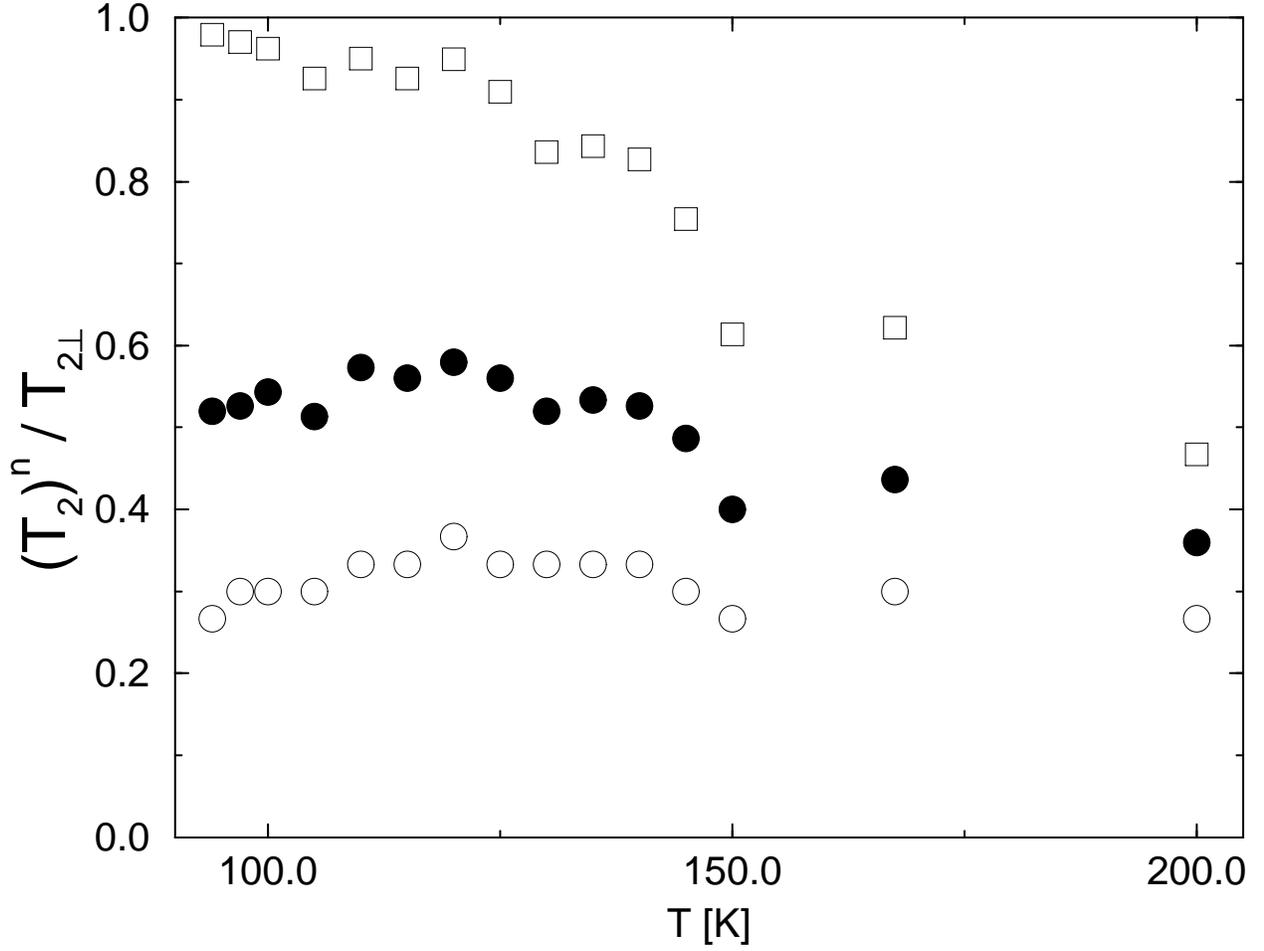}}

\protect\caption{ Experimentally determined ratio of $1/T_{2\perp}$ to nth
  power of $1/T_{2}$ for n=1 ($\Box$), 2 ($\bullet$), 3 ($\circ$) in arbitrary
  units.  That the n=3 ($\circ$) curve has less temperature dependence
  than the n=2 ($\bullet$) curve suggests
  the existence of a growing magnetic correlation length.  }

\end{figure}

\begin{figure}[h]

\protect\centerline{\epsfbox{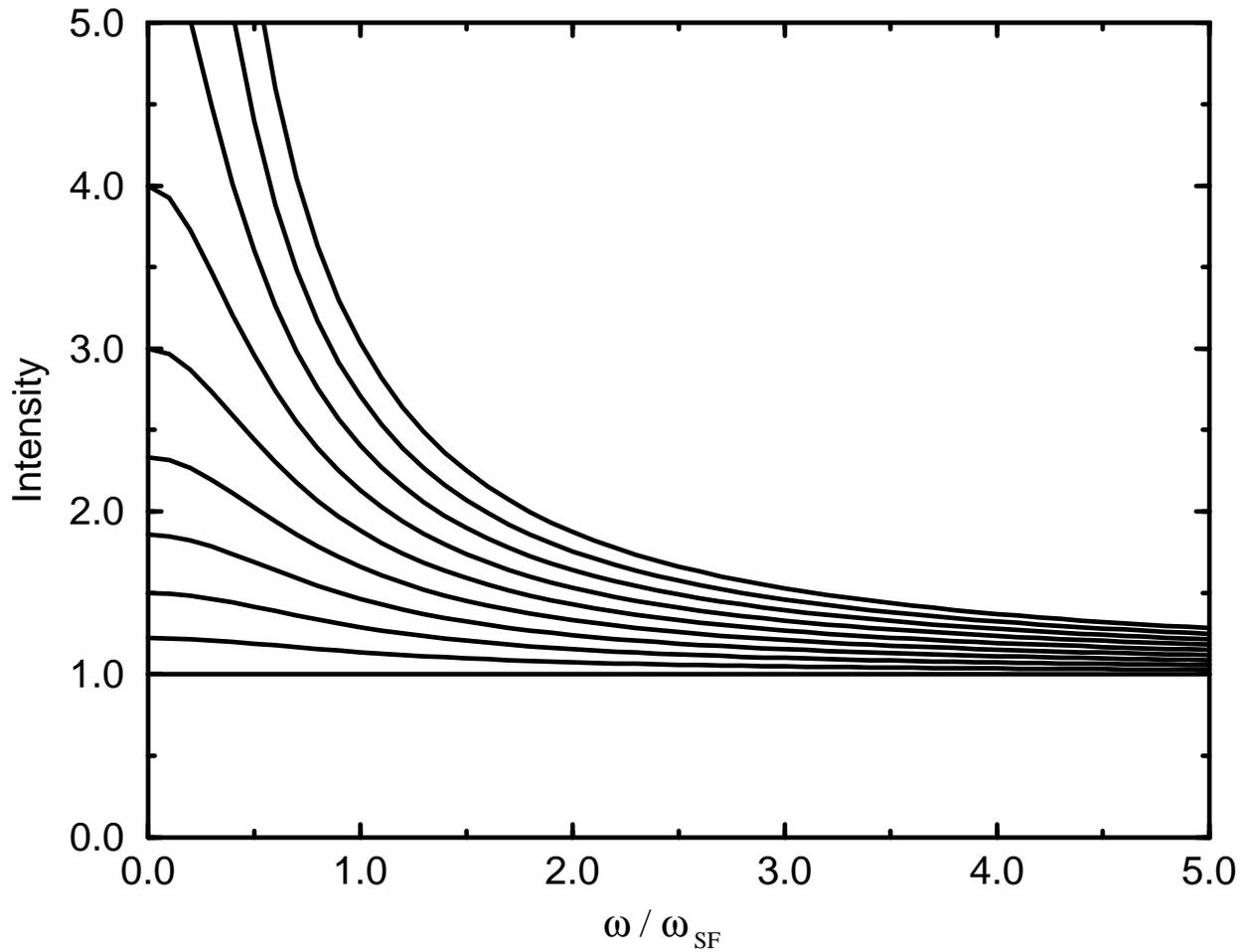}}

\protect\caption{ Calculated ratio of $q$-integrated odd-parity neutron
  absorption to q-integrated even parity neutron absorption, plotted versus
  frequency for $\Delta = J_{\perp}\chi_{max}$ = 0.0, 0.1 \ldots 0.9.
  $\Delta=0.0$ corresponds to the lowest curve and $\Delta=0.9$ to the top
  curve.}

\end{figure}

\end{document}